\documentclass{elsart5p}

\usepackage{graphicx}
\usepackage{amssymb}
\usepackage{dcolumn,booktabs}

\newcolumntype{z}[1]{D{.}{.}{#1}}

\begin{document}

\begin{frontmatter}

% Title, authors and addresses

% use the thanksref command within \title, \author or \address for footnotes;
% use the corauthref command within \author for corresponding author footnotes;
% use the ead command for the email address,
% and the form \ead[url] for the home page:
% \title{Title\thanksref{label1}}
% \thanks[label1]{}
% \author{Name\corauthref{cor1}\thanksref{label2}}
% \ead{email address}
% \ead[url]{home page}
% \thanks[label2]{}
% \corauth[cor1]{}
% \address{Address\thanksref{label3}}
% \thanks[label3]{}

\title{Low temperature electron dephasing time in AuPd revisited}

% use optional labels to link authors explicitly to addresses:
% \author[label1,label2]{}
% \address[label1]{}
% \address[label2]{}

\author[label1,label2]{J. J. Lin,\corauthref{cor1}}
\corauth[cor1]{Corresponding author. Institute of Physics and Department of Electrophysics,
National Chiao Tung University, Hsinchu 30010, Taiwan.}
\ead{jjlin@mail.nctu.edu.tw}
\author[label2]{T. C. Lee,}
\author[label1]{and S. W. Wang}

\address[label1]{Institute of Physics, National Chiao Tung University, Hsinchu 30010, Taiwan}
\address[label2]{Department of Electrophysics, National Chiao Tung University, Hsinchu 30010, Taiwan}

\begin{abstract}

Ever since the first discoveries of the quantum-interference transport in mesoscopic systems, the
electron dephasing times, $\tau_\varphi$, in the concentrated AuPd alloys have been extensively
measured. The samples were made from different sources with different compositions, prepared by
different deposition methods, and various geometries (1D narrow wires, 2D thin films, and 3D thick
films) were studied. Surprisingly, the low-temperature behavior of $\tau_\varphi$ inferred by
different groups over two decades reveals a systematic correlation with the level of disorder of
the sample. At low temperatures, where $\tau_\varphi$ is (nearly) independent of temperature, a
scaling $\tau_\varphi^{\rm max} \propto D^{-\alpha}$ is found, where $\tau_\varphi^{\rm max}$ is
the maximum value of $\tau_\varphi$ measured in the experiment, $D$ is the electron diffusion
constant, and the exponent $\alpha$ is close to or slightly larger than 1. We address this
nontrivial scaling behavior and suggest that the most possible origin for this unusual dephasing
is due to dynamical structure defects, while other theoretical explanations may not be totally
ruled out.

\end{abstract}


\end{frontmatter}

% main text
\section{Introduction}
\label{}
% The Appendices part is started with the command \appendix;
% appendix sections are then done as normal sections
% \appendix

% \section{}
% \label{}

The electron dephasing time, $\tau_\varphi$, in mesoscopic structures at very low temperatures has
recently attracted intense theoretical \cite{Zaikin98,Zawa99,Imry99} and experimental
\cite{Mohanty97,Natelson01,Pierre03,Bauerle03,Hackens05} attention. One of the key issues under
exquisite discussion is whether $\tau_\varphi$ should diverge to an infinite value or ``saturate"
to a finite value as $T \rightarrow 0$ K. It is known that even very dilute magnetic scattering,
if any exists, could eventually dominate over all other kinds of inelastic electron scattering (in
particular, the Nyquist electron-electron scattering \cite{Alt85EEI} and the electron-phonon
scattering \cite{Lin06PRB}) and control $\tau_\varphi$ at sufficiently low temperatures, leading
to a seemingly saturated behavior over a certain range of temperature. The magnetic scattering
time, $\tau_m$, in clean metals containing dilute magnetic impurities have very recently been
extensively checked theoretically \cite{Zarand04,Micklitz06,Kettemann} and experimentally
\cite{Bauerle03,Mallet,Alzoubi}. On the other hand, it has been argued if intrinsic
electron-electron interactions in the presence of disorder might lead to saturation in
$\tau_\varphi (T \rightarrow 0)$ \cite{Zaikin98}. There are also theories proposing that materials
properties associated with specific dynamical structural defects (often modeled as two-level
tunneling systems \cite{Zawa05}) may cause noticeable dephasing at low temperatures
\cite{Zawa99,Imry99}. In spite of these different opinions, one consensus has been reached by
several groups, saying that the responsible electron dephasing processes in {\em highly
disordered} and {\em weakly disordered} metals might be {\em dissimilar}
\cite{Pierre03,Zawa05,Ovadyahu01}. That is, while magnetic scattering is responsible in weakly
disordered metals \cite{Zarand04,Glazman03}, a different mechanism may be relevant for the
``saturation" or very weak temperature dependence of $\tau_\varphi$ found in highly disordered
alloys \cite{Lin02EPL}. Thus, systematic studies of the temperature and disorder behaviors of
$\tau_\varphi$ in highly disordered metals should be of great interest, and would provide valuable
information complementary to that learned from weakly disordered metals
\cite{Mohanty97,Pierre03,Bauerle03,Mallet,Alzoubi}.

For the convenience of discussion, those samples recently studied in Refs. \cite{Mohanty97},
\cite{Pierre03}, \cite{Bauerle03}, \cite{Mallet} and \cite{Alzoubi}, where the values of diffusion
constant $D$ are comparatively large (typically, $D \gtrsim$ 100 cm$^2$/s), will be referred to as
being ``weakly disordered," while the samples with values of $D \sim$ 10 cm$^2$/s and smaller to
be discussed in this work will be referred to as being ``highly disordered." In the free-electron
model, $D = v_F^2 \tau/d = (\hbar /md)(k_Fl)$, where $v_F$ ($k_F$) is the Fermi velocity (wave
number), $m$ is the effective electron mass, $\tau$ ($l$) is the electron elastic mean free time
(path), and $d$ is the dimensionality of the sample. That is, in the following discussion we
assume $D \propto k_Fl \propto \rho^{-1}$, where $\rho$ is the (residual) resistivity.

We shall survey and discuss the low-temperature electron dephasing times in the concentrated
gold-palladium (AuPd) alloys whose resistances, magnetoresistances and $\tau_\varphi$ have
previously been extensively measured by several groups
\cite{Natelson01,Lin02EPL,Lin98PRL,Lin87ng2,Webb98,Trion04,Lin87ng1,Heraud87,Chang05}. The reasons
for our revisiting the temperature and disorder behaviors of $\tau_\varphi$ of this particular
material are discussed below. Here we emphasize that the electron dephasing time $\tau_\varphi =
\tau_\varphi (T,D)$ or $\tau_\varphi = \tau_\varphi (T,\rho)$ is both a function of temperature
and a function of disorder.

\begin{enumerate}
\item Ever since the first works of Dolan and Osheroff \cite{Dolan79}, and Giordano {\it et al.}
\cite{ng79} on the weak-localization and electron-electron interaction effects in 1979 in AuPd,
this material has been continuously studied by several groups until nowadays. Narrow wires, thin
films, and thick films have been fabricated and studied, corresponding to one- (1D), two- (2D) and
three-dimensional (3D) systems, respectively, with regards to the weak-localization
\cite{Bergmann84} and electron-electron interaction \cite{Alt85EEI} effects. Thus far, there
already exist in the literature a good amount of data based on this single material. Therefore, a
close examination of any possible correlation among those {\em independently measured} dephasing
times $\tau_\varphi$ should be desirable and insightful.

\item In the course of the quantum-interference studies, the source materials used by different
groups were obtained from different suppliers, and the molar compositions Au$_{100-x}$Pd$_x$ used
were different. In most cases, $x$ varied between 40 and 60 and, thus, the samples fell in the
concentrated alloy regime. Moreover, different techniques were employed for the fabrication of
samples, including the thermal-flash evaporation, electron-beam evaporation, DC sputtering, and RF
sputtering deposition methods. Different substrates such as glass, quartz, and sapphire substrates
were also adopted in different experiments; whereas it is known that quartz and sapphire
substrates contain far fewer (magnetic) impurities than glass substrates do.

As a consequence, the levels of magnetic impurities contained in the bulk of the samples, if
exist, should be different from sample to sample prepared by different groups. Furthermore, in the
case of narrow wires and thin films, the amounts of magnetic scattering due to the surfaces and
the interfaces between the sample and the substrate, if exist, should also differ from sample to
sample prepared by different groups. Therefore, one should expect {\em randomly distributed}
values of the ``saturated" dephasing time $\tau_\varphi^{\rm max}$ for the various samples, if the
measured $\tau_\varphi^{\rm max}$ should originate from the magnetic scattering due to {\em
unintentional} magnetic impurity contamination. Notice that, in this work, we denote the highest
value of $\tau_\varphi$ extracted at the lowest measurement temperature in each experiment as
$\tau_\varphi^{\rm max}$.

\item The measured low-temperature $\tau_\varphi$ in essentially all 1D, 2D and 3D AuPd samples
revealed a very weak temperature dependent behavior already at a {\em relatively high} temperature
of around 1$-$4 K. That is, for reasons yet to be fully understood, the ``saturation" behavior of
$\tau_\varphi$ is particularly strong in this material. Therefore, close examination of the
properties of $\tau_\varphi$ in this {\em highly disordered} alloy material may shed light on our
understanding of the puzzling ``saturation problem" \cite{Mohanty97,Lin02JPCM}.

\end{enumerate}

\begin{table}
\caption{Relevant parameters for the AuPd (AgPd) samples whose highest measured values of the
electron dephasing time $\tau_\varphi^{\rm max}$ are collected in Fig. \ref{f1}. The maximum
electron dephasing length $L_\varphi^{\rm max} = (D \tau_\varphi^{\rm max})^{1/2}$. The values of
electron diffusion constant were computed using the 3D form $D = v_Fl/3$ for all narrow wire, thin
film and thick film samples. For sample groups A to J, the parameters are taken from Refs.
\cite{Lin98PRL}, \cite{Lin02EPL}, \cite{Lin01JPCM}, \cite{Lin87ng2}, \cite{Webb98},
\cite{Natelson01}, \cite{Trion04}, \cite{Lin87ng1}, \cite{Heraud87} and \cite{Chang05},
respectively. All sample groups indicate AuPd alloys except the sample group C which indicates
AgPd alloys. In the last column, the letter ``y" indicates that the experimental
$\tau_\varphi^{\rm max}$ is already saturated, the letter ``w" indicates that the experimental
$\tau_\varphi^{\rm max}$ already reveals a much weaker temperature dependence than theoretically
expected, the letter ``n" indicates no saturation at the lowest measurement temperature in that
particular experiment, and the symbol ``$-$" indicates that the temperature behavior of
$\tau_\varphi$ was not demonstrated in the cited reference.} \label{tab1}

\renewcommand{\arraystretch}{.9}
%\begin{tabular}{lp{0.9cm}p{1.1cm}p{1.2cm}p{1.3cm}c}
\begin{tabular}{lp{9mm}cccc}
\toprule Sample Group & $l$(nm) & $L_{\phi}^{\max}$(nm)&
$D$(cm$^{2}$/s) & $\tau _{\varphi }^{\max}$(ps)& Saturation\\
\midrule A (3D-AuPd) & 0.57 & 58 & 2.65 & 12.7 & y \\
& 0.76 & 69 & 3.54 & 13.5 & y \\
& 1.87 & 56 & 8.72 & 3.56 & y \\
& 1.00 & 68 & 4.66 & 10 & y \\
& 1.82 & 74 & 8.49 & 6.40 & y \\
& 1.14 & 85 & 5.31 & 13.5 & y \\
B (3D-AuPd) & 1.05 & 94 & 4.9 & 18 & y \\
& 0.28 & 105 & 1.3 & 85 & y \\
& 0.28 & 107 & 1.3 & 88 & y \\
& 0.74 & 86 & 5.3 & 14 & y \\
C (3D-AgPd) & 0.18 & 88 & 0.85 & 91.7 & y \\
& 0.24 & 106 & 1.1 & 103 & y \\
& 0.30 & 108 & 1.4 & 83.6 & y \\
& 0.21 & 120 & 1.0 & 145 & y \\
& 0.12 & 116 & 0.57 & 236 & y \\
& 0.21 & 116 & 1.0 & 135 & y \\
& 0.096 & 139 & 0.45 & 430 & y \\
& 0.075 & 110 & 0.35 & 347 & y \\
& 0.118 & 141 & 0.55 & 362 & y \\
& 0.092 & 69 & 0.43 & 111 & y \\
D (2D-AuPd) & 0.47 & 42 & 2.2 & 8 & y \\
& 0.47 & 81 & 2.2 & 30 & y \\
& 0.47 & 65 & 2.2 & 19 & y \\
E (2D-AuPd) & 0.17 & 62 & 0.8 & 48.5 & y \\
F (1\&2D-AuPd) & 2.57 & 147 & 12 & 18.1 & n \\
& 2.57 & 108 & 12 & 9.8 & n \\
& 3.21 & 112 & 15 & 8.3 & n \\
& 3.21 & 130 & 15 & 11.3 & n \\
& 3.21 & 115 & 15 & 8.8 & n \\
& 3.21 & 147 & 15 & 14.4 & w \\
& 3.21 & 176 & 15 & 20.7 & w \\
& 3.21 & 151 & 15 & 15.3 & w \\
& 3.21 & 139 & 15 & 12.8 & w \\
& 3.21 & 143 & 15 & 13.6 & w \\
G (1\&2D-AuPd) & 2.87 & 190 & 13.4 & 26.94 & y \\
& 2.87 & 180 & 13.4 & 24.18 & y \\
& 1.69 & 150 & 7.9 & 28.48 & y \\
H (1D-AuPd) & 4.9 & 80 & 23 & 2.8 & y \\
& 4.9 & 61 & 23 & 1.6 & y \\
& 4.9 & 96 & 23 & 4 & y \\
I (1D-AuPd) & 4.8 & 98 & 19.2 & 5 & y \\
J (1D-AuPd) & 1.5 & 27 & 7 & 1.04 & $-$ \\
& 1.2 & 38 & 5.6 & 2.58 & $-$ \\
& 2.5 & 81 & 12 & 5.47 & $-$ \\
\bottomrule
\end{tabular}
\renewcommand{\arraystretch}{1}

\end{table}

In this paper, we have collected 34 data points of $\tau_\varphi^{\rm max}$ measured on AuPd
samples from the literature. Those values of $\tau_\varphi^{\rm max}$ were all extracted from the
weak-localization magnetoresistance measurements, with only one exception in Ref. \cite{Trion04}
where the value of $\tau_\varphi^{\rm max}$ was determined from the time-dependent universal
conductance fluctuation measurements. A set of the $\tau_\varphi^{\rm max}$ data (10 data points)
measured on a series of Ag$_{40}$Pd$_{60}$ samples are also included in the present discussion,
because AgPd and AuPd have very similar materials and electronic properties \cite{Hansen58}. The
relevant parameters for the samples surveyed in this work are listed in Table \ref{tab1}. In
addition, we have measured the low temperature thermoelectric powers (the Seebeck coefficient) in
both as-prepared and thermally annealed bulk AuPd. We have also fabricated thin and thick AuPd
films and measured their resistances in zero field and in (moderately) high magnetic fields down
to subkelvin temperatures. These thermoelectric power and resistivity measurements provide useful
auxiliary information about the possible existence or not of localized magnetic moments in this
alloy material.

\section{Results and discussion}

\subsection{Electron dephasing time}

The electron dephasing times $\tau_\varphi$ in AuPd narrow wires
\cite{Natelson01,Lin87ng2,Trion04,Heraud87,Chang05}, thin films
\cite{Natelson01,Webb98,Trion04,Lin87ng1}, and thick films \cite{Lin02EPL,Lin98PRL} have been
extensively measured over the past two decades. In general, a temperature dependent dephasing time
is observed at a few degrees Kelvin and higher, which can be satisfactorily attributed to the
Nyquist electron-electron scattering and/or the electron-phonon scattering. Between about 1 and 4
K, depending on samples, a crossover to a very weak temperature dependence or a seemingly
saturation of $\tau_\varphi$ is found (for example, see Fig. 1 of Ref. \cite{Lin87ng2}, Fig. 1 of
Ref. \cite{Lin02EPL}, and Fig. 4 of Ref. \cite{Trion04}). Such a ``crossover temperature" of 1--4
K is significantly higher than those recently observed in the weakly disordered metals studied in,
for example, Refs. \cite{Mohanty97}, \cite{Pierre03}, \cite{Bauerle03} and \cite{Hackens05}.

\begin{figure}[t]\label{fig1}
\includegraphics[scale=1]{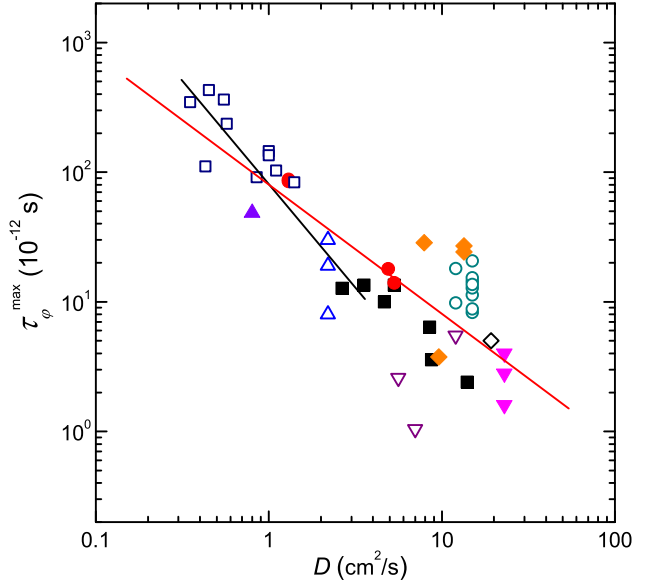}
\caption{\label{f1} (Color online) Variation of $\tau_\varphi^{\rm max}$ with diffusion constant
$D$ for concentrated AuPd and AgPd alloys. The measured $\tau_\varphi^{\rm max}$ are taken from:
closed squares (Ref. \cite{Lin98PRL}), closed circles (Ref. \cite{Lin02EPL}), open squares (Ref.
\cite{Lin01JPCM}), open up triangles (Ref. \cite{Lin87ng2}), closed up triangles (Ref.
\cite{Webb98}), open circles (Ref. \cite{Natelson01}), closed diamonds (Ref. \cite{Trion04}),
closed down triangles (Ref. \cite{Lin87ng1}), open diamonds (Ref. \cite{Heraud87}), and open down
triangles (Ref. \cite{Chang05}). The long straight line indicates the empirical scaling
$\tau_\varphi^{\rm max} \propto D^{-1}$ given by Eq.~(\ref{eq1}). The short straight line drawn
between $D \approx$ 0.3 and 3 cm$^2$/s indicates the scaling $\tau_\varphi^{\rm max} \propto
D^{-1.6}$.}
\end{figure}

Figure~\ref{f1} shows the variation of the measured value of $\tau_\varphi^{\rm max}$ with
diffusion constant $D$ collected from ten {\em independent studies} previously reported in the
literature. The symbols are the experimental data, as indicated in the caption to Fig.~\ref{f1}.
Noticeably, inspection of Fig.~\ref{f1} reveals that the measured values of $\tau_\varphi^{\rm
max}$ are {\em not} randomly distributed, as one might have naively expected for
magnetic-scattering-induced dephasing. Needless to say, if the dephasing were due to {\em
arbitrary contaminations} of magnetic impurities in the various samples made by different groups
over different times, one should have observed {\em randomly distributed} values of
$\tau_\varphi^{\max}$. On the contrary, we find that {\em there is a strong correlation among the
experimental values of $\tau_\varphi^{\rm max}$ with the levels of disorder contained in the
samples}, independent of how and where the samples were made. In Fig.~\ref{f1}, the long straight
solid line is drawn to guide the eye and is given by
\begin{equation}
\tau_\varphi^{\rm max} \approx 0.08\, D^{-1} \,\,\,\, \rm{[ns]}\,, \label{eq1}
\end{equation}
where $D$ is in cm$^2$/s. Figure~\ref{f1} suggests an approximate empirical scaling
$\tau_\varphi^{\rm max} \propto D^{-1}$, which holds for over two decades of the $D$ value from
about 0.3 to 30 cm$^2$/s, corresponding to $\tau_\varphi^{\rm max}$ varying roughly from $\sim 3
\times 10^{-10}$ to $\sim 3 \times 10^{-12}$ s. Equivalently, $\tau_\varphi^{\max} \propto
(k_Fl)^{-1}$ or $\tau_\varphi^{\max} \propto \rho$ in this alloy system, and thus {\em the
``saturated" dephasing time is longer in more disordered samples}. This result is intriguing,
which should be suggestive of some unusual and yet to be understood electron dephasing mechanism
operating in this particular (and maybe also other) highly disordered materials. We notice that
Eq.~(\ref{eq1}) implies a relatively short and (almost) {\em constant} saturated dephasing length
$L_\varphi^{\rm max} = \sqrt{D \tau_\varphi^{\rm max}} \sim$ 900 $\rm \AA$ in AuPd.

We would like to mention that the values of the experimental $\tau_\varphi^{\rm max}$ considered
in Fig.~\ref{f1} and Table~\ref{tab1} are all already saturated or already reveal a much weaker
temperature dependence than theoretically expected. There is only in one case involving 5-nm wide
AuPd wires \cite{Natelson01} where no saturation in $\tau_\varphi$ was envisaged down to the
lowest measurement temperature of 80 mK in that experiment. In addition, we should point out that
it cannot be totally ruled out that the variation in Fig.~\ref{f1} may be described by the
approximation $\tau_\varphi^{\rm max} \propto D^{-\alpha}$ with an exponent $\alpha$ {\em slightly
larger than 1} \cite{Lin02JPCM,Lin01JPCM}, especially if we were to concentrate on the most
strongly disordered regime of, e.g., $D \lesssim$ 3 cm$^2$/s.

As mentioned, the AuPd samples shown in Fig.~\ref{f1} are either narrow wires, thin films, or
thick films, which are 1D, 2D, or 3D with respect to the weak-localization effects. Nevertheless,
with regard to the classical Boltzmann transport (the Drude conductivity $\rho^{-1}$), all the
samples studied are 3D, because the elastic electron mean free path $l$ is short in AuPd, as
compared with the narrow wire diameter or the film thickness. (For the alloys collected in
Fig.~\ref{f1}, the value of $l$ varies approximately between 2 and 49 \r{A}, see
Table~\ref{tab1}.) Therefore, it is justified to discuss the measured $\tau_\varphi^{\rm max}$ for
all the narrow wire, thin film and thick film samples on an equal footing in terms of the 3D form
of the electron diffusion constant $D = v_Fl/3$.

It is worth noting that a reminiscent scaling $\tau_\varphi^{\rm max} \propto D^{-1}$ for a good
number of 3D {\em polycrystalline} metal alloys has previously been reported by Lin and Kao
\cite{Lin01JPCM}. A related discussion on the correlation between $\tau_\varphi$ at a given
temperature and the level of disorder $(k_Fl)^{-1}$ has been reported for a series of 3D
In$_2$O$_{3-x}$ thick films \cite{Ovadyahu84}, and a dephasing time $\tau_\varphi ({\rm 4.2\,K})
\propto \rho$ has been reported for a series of 2D In$_2$O$_{3-x}$ thin films \cite{Ovadyahu83} by
Ovadyahu. (However, it should be noted that in Refs. \cite{Ovadyahu84} and \cite{Ovadyahu83}, the
disorder dependent electron dephasing time was inferred for a given temperature where
$\tau_\varphi$ is still governed by a {\em strongly temperature dependent} dephasing mechanism.)
Very recently, in a series of 2D Cu$_{93}$Ge$_4$Au$_3$ thin films, Huang {\it et al.}
\cite{Lin07PRL} have observed  a dephasing time which first ``saturates" around 6 K and then
crosses over to a slow increase with further decrease in temperature. At 0.4 K, an approximate
scaling $\tau_\varphi ({\rm 0.4\,K}) \propto R_\square$ is found, where $R_\square$ is the sheet
resistance. These above-mentioned results seem to suggest that, in certain metals and alloys,
strong electron dephasing may originate from specific structural defects in the samples
\cite{Imry03,Lin07PRL}. We should also notice that the authors of Ref. \cite{Imry03} have
presented measurements that particularly argued against magnetic scattering as a cause for the
``saturation" in $\tau_\varphi (T \rightarrow 0)$.

\subsection{Low temperature thermoelectric powers}

The thermoelectric power (thermopower), $S$, is a quantity which is known to be extremely
sensitive to the existence of a trace amount of magnetic impurities in an otherwise pure metal
\cite{MacDonald62}. In typical pure nonmagnetic metals, the low temperature thermopower is
comprised of two terms: $S = AT + BT^3$, where the first and the second terms represent
contributions from electron diffusion and phonon drag, respectively. At temperatures of a few
degrees Kelvin, the linear term usually dominates. However, in the presence of a small amount (for
instance, $\sim$ a few tenths or a few ppm) of magnetic impurities, a {\em very large} $S$ is
found (typically, reaching $\sim$ several $\pm \mu$V/K), which is well more than an order of
magnitude larger than that in the corresponding pure metal. In addition, the thermopower now
reveals a {\em broad} maximum, resulting in a nearly constant $S$ over a wide range of temperature
at liquid-helium temperatures. For example, the value of $S$(4\,K) changes from +0.03 $\mu$V/K for
pure Au to $-$7.2 $\mu$V/K for the AuFe Kondo alloy containing 13 ppm of Fe \cite{Kopp75}. In the
case of the Au$_{100-x}$Pd$_x$ alloys, the low temperature thermopowers have been extensively
measured by Rowland {\it et al.}, \cite{Rowland74} and Gu$\acute{e}$nault {\it et al.}
\cite{Guenault78} over the whole alloy series. They pointed out that the thermopower can be well
described by the expression $S = AT + BT^3 + CT/(T+0.2)$, in which the third term represents an
extra contribution from Fe contamination. Notably, they observed that a finite value of $C$ could
only be inferred for the Au-rich ($x \lesssim 20$) alloys; whereas in the concentrated alloys with
the compositions ($x \approx$ 40$-$60) pertinent to our discussion, they found $C = 0$, i.e., no
magnetic impurities could be inferred from their thermopower measurements. Their studies seem to
indicate that Fe atoms may {\em not} readily form localized magnetic moments in the concentrated
AuPd alloys. This issue deserves further investigation.

\begin{figure}[t]\label{fig2}
\includegraphics[scale=1]{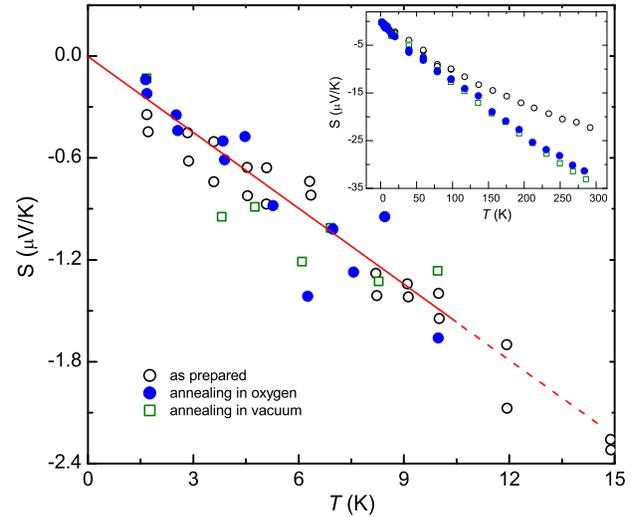}
\caption{\label{f2} (Color online) Thermoelectric power as a function of temperature for the
as-prepared ($\circ$), the oxygen annealed ($\bullet$) and the vacuum annealed ($\Box$) bulk
Au$_{60}$Pd$_{40}$ (0.5-mm diameter and $\sim$ 1 cm long) between 1.5 and 15 K. The straight line
is a least-squares fit to the data between 1.5 and 10 K. Inset: Thermopower as a function of
temperature between 1.5 and 300 K for the same samples.}
\end{figure}

In this work, we used two pieces of {\em bulk} Au$_{60}$Pd$_{40}$ (0.5-mm diameter and $\sim$ 1 cm
long, and 99.98\% purity) for thermopower measurements between 1.5 and 300 K to check whether
magnetic impurities might play important roles in concentrated AuPd alloys. The first bulk
Au$_{60}$Pd$_{40}$ was measured twice, once before and once after a thermal annealing at
800$^{\circ}$C for 16 h in an oxygen atmosphere of $\approx 5 \times 10^{-3}$ torr
\cite{Barnard73}. The second bulk sample was first thermally annealed at 800$^{\circ }$C for 16 h
in a vacuum of $\approx 5 \times 10^{-4}$ torr before its thermopower was measured.
Figure~\ref{f2} shows the variation of thermopower with temperature for our two bulk AuPd samples,
as indicated. To within our experimental uncertainties, the absolute value of the thermopower does
{\em not at all} decrease after the thermal annealing in oxygen. The thermal annealing in vacuum
also does not change the magnitude and the temperature behavior of the thermopower. This result
demonstrates that the amount of magnetic (e.g., Fe) impurities, if any exists, in the concentrated
AuPd alloys is negligibly small. Otherwise, the magnetic impurities should have become oxidized
after annealing, losing their moments, and their contributions to $S$ greatly suppressed. Between
1 and 10 K, our measured thermopower can be well described by the linear expression $S = - 0.14 T$
$\mu$V/K, as indicated by the least-squares fitted straight solid line in Fig.~\ref{f2}. This
result is in consistency with that previously reported by Gu$\acute{e}$nault {\it et al.}
\cite{Guenault78}. In short, no evidence of a huge thermopower with a broad maximum at a few
degrees Kelvin signifying the existence of an appreciable level of magnetic impurities is found in
the concentrated AuPd alloys. Indeed, to the best of our knowledge, we are aware of no report on
the AuPd alloy being a Kondo system. The inset of Fig.~\ref{f2} shows a plot of the overall
temperature behavior of $S$ between 1.5 and 300 K for our as-prepared and annealed bulk AuPd
samples.

\subsection{Resistances in zero field and in magnetic fields}

\begin{figure}[t]\label{fig3}
\center
\includegraphics[scale=0.58]{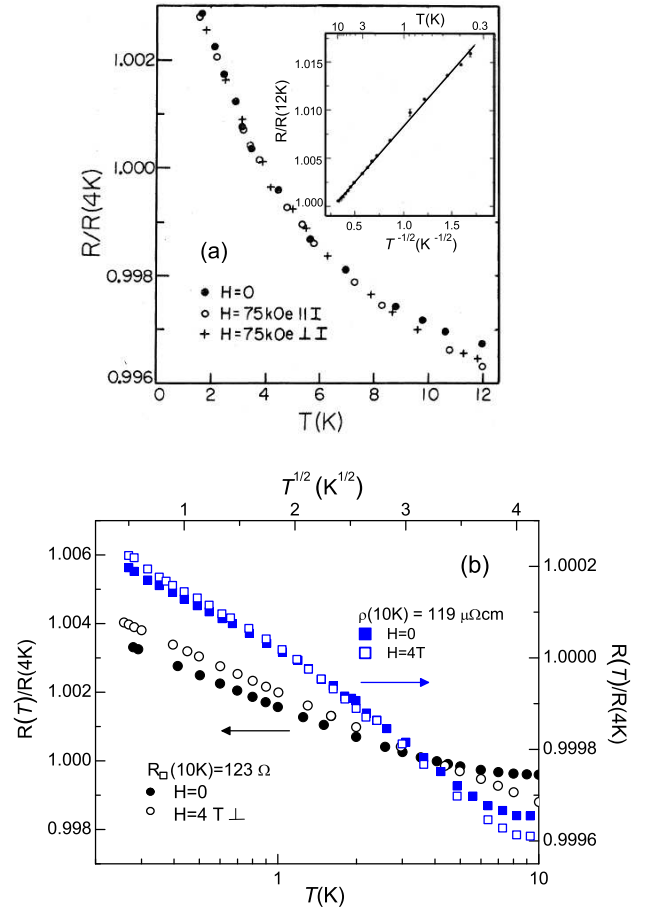}
\caption{\label{f3}(Color online) (a) Resistance as a function of temperature for a 40-nm diameter
AuPd narrow wire in zero magnetic field ($\bullet$) and in a perpendicular magnetic field of 7.5 T
($\circ$). The inset shows the normalized resistance $R(T)/R$(12\,K) as a function of $1/\sqrt{T}$
(taken from Ref. \cite{ng80}). (b) Resistance as a function of the logarithm of temperature for a
40-nm thick AuPd film in zero field ($\bullet$) and in a perpendicular magnetic field of 4 T
($\circ$). Also shown is the resistance as a function of the square root of temperature for a
560-nm thick AuPd film in zero field ($\blacksquare$) and in a perpendicular magnetic field of 4 T
($\square$).}
\end{figure}

To investigate whether the observed weak temperature dependence or ``saturation" of
$\tau_\varphi^{\rm max}$ in the AuPd samples might be due to magnetic scattering, it is
instructive to examine the temperature behavior of resistance in both zero field and in
(moderately) high magnetic fields. In this work we have fabricated thin (2D) and thick (3D)
Au$_{50}$Pd$_{50}$ films for low temperature resistance measurements. Our films were made by DC
sputtering deposition on glass substrates held at ambient temperature. A background pressure of $9
\times 10^{-7}$ torr was reached before the sputtering was initiated. An argon atmosphere of
20--30 mtorr was maintained during the sputtering process. The resistances as a function of
temperature were measured down to 0.3 K. Figures~\ref{f3}(a) and \ref{f3}(b) show the variations
of resistance with temperature for a AuPd narrow wire (taken from Ref. \cite{ng80}), a thin film
and a thick film, respectively, in zero field and in a perpendicular magnetic field, as indicated.
These figures indicate that the resistance rises vary with $- 1/\sqrt{T}$ and $-$ln$T$ all the way
down to 0.3 K in narrow wires and thin films, respectively, as would be expected from the
electron-electron interaction effects in the presence of disorder \cite{Alt85EEI}. In the case of
thick films, the electron-electron interaction effects are comparatively small and the expected $-
\sqrt{T}$ dependence \cite{Alt85EEI} is somewhat masked by the intrinsic temperature behavior of
resistivity (due to scattering from localized exchange-enhanced Pd $d$ states \cite{Murani74}) of
this alloy material. Most importantly, Figs.~\ref{f3}(a) and \ref{f3}(b) indicate that, in all
three geometries in zero magnetic field, there is no any sign of a crossover to a resistivity
saturation down to 0.3 K, as would be expected for the Kondo effect \cite{Mallet,Daybell67}, even
though a very weak temperature dependence of $\tau_\varphi$ is already set in at temperatures
($\sim$ 1$-$4 K) about an order of magnitude higher. Moreover, in all geometries in the presence
of a magnetic field of a few T, there is no any evidence of a detectable negative
magnetoresistance signifying the alignment effect of localized magnetic spins \cite{Monod67}.
Thus, both the resistance and magnetoresistance behaviors do not seem to suggest the existence of
noticeable localized magnetic moments in this alloy material. In fact, a similar conclusion of no
Kondo effect in AuPd has previously been drawn by Giordano \cite{ng80} from comparison of the
temperature behavior of resistance in as-sputtered and annealed films.

\section{Summary and conclusion}

In this work, we survey the low-temperature electron dephasing times in the AuPd (and AgPd) alloys
measured by several groups over the past 20 years. We show the existence of a nontrivial scaling
$\tau_\varphi^{\rm max} \propto D^{-\alpha}$, with $\alpha$ close to or slightly larger than 1,
among the various samples made of this alloy material. This intriguing scaling behavior strongly
suggests that the observed $\tau_\varphi^{\rm max}$ in AuPd cannot be simply attributed to the
magnetic scattering time arising from random amounts of magnetic impurity contamination. To within
our experimental uncertainties, the thermopower measurements at liquid-helium temperatures, and
the resistance measurements in both zero field and in a (moderately) high magnetic field, indicate
no sign of the Kondo effect in this concentrated alloy system. As a matter of fact, it should be
noted that previous measurements on as-prepared and then annealed AuPd thin films \cite{Lin87ng2}
and thick films \cite{Lin02EPL} have also ruled out the magnetic scattering as a plausible
explanation for the ``saturation" of $\tau_\varphi$ found in this material \cite{Lin03JPS}. In
addition, previous studies of AuPd {\em thin} films deposited on both glass and quartz substrates
indicated no difference in the temperature behavior of $\tau_\varphi$, although it was thought
that glass substrates should contain more (magnetic) impurities which should in turn contribute to
dephasing through electron scattering at the film-substrate interface \cite{Lin87ng2}.

Theoretically, for {\em highly disordered} 3D systems, a $\tau_\varphi$ possessing a very weak
temperature dependence in a certain temperature interval and then crossing over to a slow increase
with decreasing temperature has recently been predicted in a model based on tunneling states of
dynamical structural defects \cite{Galperin04}. This model also predicts a `counterintuitive'
scaling $\tau_\varphi \propto D^{-1}$ in the plateau-like region. Our observation in Fig. \ref{f1}
essentially mimic these {\em qualitative} features. Close comparison between the experiment and
theory would require quantitative calculations using realistic material parameters
\cite{Zawa05,Zarand05}. Finally, we notice that in the weakly disordered regime, the
electron-electron interaction theory \cite{Zaikin98} predicts a higher saturation value of
$\tau_\varphi$ in cleaner samples. How this prediction might be modified in the highly disordered
regime would be of great interest. Experimentally, in their systematic studies of numerous {\em
high-mobility} GaAs/AlGaAs quantum wires, Noguchi {\it et al.} \cite{Noguchi96} have observed a
saturated dephasing time which scales approximately with mobility, i.e., $\tau_\varphi^{\rm max}
\propto \mu$, which suggests in turn a {\em direct proportionality} to $D$. This experimental
$\tau_\varphi^{\rm max}$ behavior versus disorder is in line with the prediction of the
electron-electron interaction theory \cite{Zaikin98} in the {\em weakly disordered} regime.

To summarize, our result indicates that the intriguing electron dephasing found in the AuPd alloys
is very unlikely due to magnetic scattering. It may originate from specific dynamical structure
defects in the samples. Other theoretical explanations may also be explored.

{\it Note added}: After the submission of this paper, we have learned that, very recently, Golubev
and Zaikin \cite{Zaikin07} have developed a universal formula for the saturated electron dephasing
time $\tau_\varphi^{\rm max}$ based on a model treating the effect of electron-electron
interactions on weak localization in arbitrary arrays of quantum dots. They found that
electron-electron interactions always cause a saturated dephasing time, and the saturated value of
$\tau_\varphi^{\rm max}$ depends strongly and non-monotonously on the level of disorder in the
sample while being insensitive to the sample dimensionality. Their theory has considered the
electron dephasing times in all three cases of weakly disordered conductors, strongly disordered
conductors and metallic quantum dots in a unified manner. Most notably, their theory predicts
that, in {\em weakly disordered} systems, $\tau_\varphi^{\rm max}$ increases with {\em decreasing}
disorder; while, on the contrary, in {\em strongly disordered} systems, $\tau_\varphi^{\rm max}$
increases with {\em increasing} disorder. Our observation in Fig.~\ref{f1} is in line with this
new theoretical prediction (see the discussion and the Fig. 6 in \cite{Zaikin07}). In particular,
if we focus on the strongly disordered regime with $D \approx$ 0.3$-$3 cm$^2$/s, the data in
Fig.~\ref{f1} would be best described by $\tau_\varphi^{\rm max} \approx 0.084 D^{-1.6}$ ns, as
indicated by the short straight line.

\section{Acknowledgments}

The authors are grateful to Y. Galperin, A. Zaikin and A. Zawadowski for valuable discussions.
This work was supported by the Taiwan National Science Council through Grant Nos. NSC
94-2112-M-009-035 and NSC 95-2112-M-009-013, and by the MOE ATU Program.


\begin{thebibliography}{99}

% \bibitem{label}
% Text of bibliographic item

% notes:
% \bibitem{label} \note

% subbibitems:
% \begin{subbibitems}{label}
% \bibitem{label1}
% \bibitem{label2}
% If there is a note, it should come last:
% \bibitem{label3} \note
% \end{subbibitems}

\bibitem{Zaikin98} D. S. Golubev, A. D. Zaikin, Phys. Rev. Lett. 81 (1998) 1074.

\bibitem{Zawa99} A. Zawadowski, J. von Delft, D.C. Ralph, Phys. Rev. Lett. 83 (1999) 2632.

\bibitem{Imry99} Y. Imry, H. Fukuyama, P. Schwab, Europhys. Lett. 47 (1999) 608.

\bibitem{Mohanty97} P. Mohanty, E. M. Q. Jariwala, R. A. Webb, Phys. Rev. Lett. 78
(1997) 3366; \\P. Mohanty, R. A. Webb, Phys. Rev. Lett. 91 (2003) 066604.

\bibitem{Natelson01} D. Natelson, R. L. Willett, K. W. West, L. N. Pfeiffer, Phys. Rev. Lett. 86
(2001) 1821.

\bibitem{Pierre03} F. Pierre, N. O. Birge, Phys. Rev. Lett. 89 (2002) 206804;
\\F. Pierre, A. B. Gougam, A. Anthore, H. Pothier, D. Esteve, N. O. Birge, Phys. Rev. B 68 (2003) 085413;
\\N. O. Birge, F. Pierre, arXiv:cond-mat/0401182.

\bibitem{Bauerle03} F. Schopfer, C. B\"{a}uerle, W. Rabaud, L. Saminadayar, Phys. Rev. Lett. 90
(2003) 056801;
\\C. B\"{a}uerle, F. Mallet, F. Schopfer, D. Mailly, G. Eska, L. Saminadayar, Phys. Rev. Lett. 95
(2005) 266805.

\bibitem{Hackens05} B. Hackens, S. Faniel, C. Gustin, X. Wallart, S. Bollaert,
A. Cappy, V. Bayot, Phys. Rev. Lett. 94 (2005) 146802.

\bibitem{Alt85EEI} B. L. Altshuler, A. G. Aronov, in: Electron-Electron Interactions in Disordered
Systems, Modern Problems in Condensed Matter Science, Vol. 10, ed. A. L. Efros, M. Pollak
(North-Holland, Amsterdam, 1985) pp. 1-153.

\bibitem{Lin06PRB} L. Li, S. T. Lin, C. Dong, J. J. Lin, Phys. Rev. B 74, 172201 (2006).

\bibitem{Zarand04} G. Zar\'{a}nd, L. Borda, J. von Delft, N. Andrei, Phys. Rev. Lett. 93 (2004) 107204.

\bibitem{Micklitz06} T. Micklitz, A. Altland, T. A. Costi, A. Rosch, Phys. Rev. Lett.
96 (2006) 226601; \\T. Micklitz, T. A. Costi, A. Rosch, Phys. Rev. B 75 (2007) 054406.

\bibitem{Kettemann} S. Kettemann, E. R. Mucciolo, Phys. Rev. B 75 (2007) 184407.

\bibitem{Mallet} F. Mallet, J. Ericsson, D. Mailly, S. \"{U}nl\"{u}bayir, D.
Reuter, A. Melnikov, A. D. Wieck, T. Micklitz, A. Rosch, T. A. Costi, L. Saminadayar, C.
B\"{a}uerle, Phys. Rev. Lett. 97 (2006) 226804.

\bibitem{Alzoubi} G. M. Alzoubi, N. O. Birge, Phys. Rev. Lett. 97 (2006) 226803.

\bibitem{Zawa05} O. \'Ujs\'aghy, A. Zawadowski, J. Phys. Soc. Jpn. 74 (2005) 80.

\bibitem{Ovadyahu01} Z. Ovadyahu, Phys. Rev. B 63 (2001) 235403.

\bibitem{Glazman03} M. G. Vavilov, L. I. Glazman, Phys. Rev. B 67 (2003) 115310.

\bibitem{Lin02EPL} J. J. Lin, Y. L. Zhong, T. J. Li, Europhys. Lett. 57 (2002) 872.

\bibitem{Lin98PRL} Y. L. Zhong, J. J. Lin, Phys. Rev. Lett. 80 (1998) 588.

\bibitem{Lin87ng2} J. J. Lin, N. Giordano, Phys. Rev. B 35 (1987) 1071.

\bibitem{Webb98} R. A. Webb, P. Mohanty, E. M. Q. Jariwala, Fortschr. Phys. 46 (1998) 779.

\bibitem{Trion04} A. Trionfi, S. Lee, D. Natelson, Phys. Rev. B 70 (2004) 041304.

\bibitem{Lin87ng1} J. J. Lin, N. Giordano, Phys. Rev. B 35 (1987) 545.

\bibitem{Heraud87} A. P. Heraud, S. P. Beaumont, C. D. W. Wilkinson, P. C. Main, J. R. OwersBradley, L.
Eaves, J. Phys. C: Solid State Phys. 20 (1987) L249.

\bibitem{Chang05} F. Altomare, A. M. Chang, M. R. Melloch, Y. Hong, C. W. Tu, Appl. Phys. Lett. 86 (2005)
172501.

\bibitem{Dolan79} G. J. Dolan, D. D. Osheroff, Phys. Rev. Lett. 43 (1979) 721.

\bibitem{ng79} N. Giordano, W. Gilson, D. E. Prober, Phys. Rev. Lett. 43 (1979) 725.

\bibitem{Bergmann84} G. Bergmann, Phys. Rep. 107 (1984) 1.

\bibitem{Lin02JPCM} J. J. Lin and J. P. Bird, J. Phys.: Condens. Matter 14 (2002) R501.

\bibitem{Hansen58} M. Hansen, Constitution of Binary Alloys (McGraw-Hill, New York, 1958),
pp. 41 and 224, and references therein; \\
K. Schroder, CRC Handbook of Electrical Resistivities of Binary Metallic Alloys (CRC Press, Boca
Raton, Fla., 1983), pp. 61 and 136, and references therein.

\bibitem{Lin01JPCM} J. J. Lin, L. Y. Kao, J. Phys.: Condens. Matter 13 (2001) L119.

\bibitem{Ovadyahu84} Z. Ovadyahu, Phys. Rev. Lett. 52 (1984) 569.

\bibitem{Ovadyahu83} Z. Ovadyahu, J. Phys. C: Solid State Phys. 16 (1983) L845.

\bibitem{Imry03} Y. Imry, Z. Ovadyahu, A. Schiller, arXiv:cond-mat/0312135.

\bibitem{Lin07PRL} S. M. Huang, T. C. Lee, H. Akimoto, K. Kono, J. J. Lin, to be published.

\bibitem{MacDonald62} D. K. C. MacDonald, W. B. Pearson, I. M. Templeton, Proc. R. Soc. Lon. A 266
(1962) 161.

\bibitem{Kopp75} J. Kopp, J. Phys. F: Metal Phys. 5 (1975) 1211.

\bibitem{Rowland74} T. Rowland, N. E. Cusack, R. G. Ross, J. Phys. F: Metal Phys. 4 (1974) 2189.

\bibitem{Barnard73} R. D. Barnard, J. Phys. E: Sci. Instru. 6 (1973) 508.

\bibitem{Guenault78} A. M. Gu$\acute{e}$nault, N. S. Lawson, J. Northfield, Phil. Mag. B
38 (1978) 567.

\bibitem{ng80} N. Giordano, Phys. Rev. B 22 (1980) 5635.

\bibitem{Murani74} A. P. Murani, Phys. Rev. Lett. 33 (1974) 91.

\bibitem{Daybell67} M. D. Daybell, W. A. Steyert, Phys. Rev. Lett. 18 (1967) 398.

\bibitem{Monod67} P. Monod, Phys. Rev. Lett. 19 (1967) 1113.

\bibitem{Lin03JPS} J. J. Lin, T. J. Li, Y. L. Zhong, J. Phys. Soc. Jpn. 72 (2003) Suppl. A. 7.

\bibitem{Galperin04} Y. M. Galperin, V. I. Kozub, V. M. Vinokur, Phys. Rev. B 69 (2004) 073102; \\
V. V. Afonin, J. Bergli, Y. M. Galperin, V. L. Gurevich, V. I. Kozub, Phys. Rev. B 66 (2002)
165326.

\bibitem{Zarand05} G. Zar\'{a}nd, Phys. Rev. B 72 (2005) 245103.

\bibitem{Noguchi96} M. Noguchi, T. Ikoma, T. Odagiri, H. Sakakibara, S. N.
Wang, J. Appl. Phys. 80 (1996) 5138.

\bibitem{Zaikin07} D. S. Golubev and A. D. Zaikin, Physica E (2007), this volume.

\end{thebibliography}
\end{document}